\documentstyle[axodraw,preprint,prd,aps]{revtex}
\begin{document}
\preprint{\ \vbox{
\halign{&##\hfil\cr
       AS-ITP-2001-003\cr\cr}} } \vfil
\draft
\title{A QCD-Analysis for Radiative Decays of $\Upsilon$ into $f_2(1270)$ }
\author{J.P. Ma}
\address{Institute of Theoretical Physics,\\
Academia Sinica, \\
P.O.Box 2735, Beijing 100080, China\\
e-mail: majp@itp.ac.cn}
\maketitle

\begin{abstract}
We perform a QCD analysis for the radiative decay of a heavy
$^3S_1$ quarkonium into the tensor meson $f_2(1270)$. We make an
attempt to separate the nonperturbative effect related to the
quarkonium and that related to the tensor meson, the former is
represented by NRQCD matrix elements, while the later is
parameterized by distribution amplitudes of gluons in the tensor
meson at twist-2 level and at twist-3 level. We find that at
twist-2 level the helicity $\lambda$ of the tensor meson can be
$0$ and $2$ and the amplitude with $\lambda =2$ is suppressed.
At twist-3 level the tensor meson can have $\lambda =1$. A comparison
with experiment is made,  an agreement of our results with
experiment can be found. We also briefly discuss the radiative
decay into $\eta$ and obtain a prediction for
$\Upsilon\to\gamma+\eta$. \vskip25pt \noindent PACS numbers:
13.25.Gv, 14.40.-n, 12.38Bx.\newline Key Words: Quarkonium, Tensor
meson, Radiative decay, Factorization.
\end{abstract}

\eject
\baselineskip=15pt

\vspace{-5mm}

\vskip20pt \narrowtext
\noindent
{\bf 1. Introduction}

\vskip20pt
Decays of heavy quarkonia into light hadrons are forbidden processes by OZI rule, their
study will help us to understand how gluons, which are fundamentally dynamical
freedoms of QCD, are converted into light hadrons. Radiative decays of quarkonia
into a single hadron
provide an ideal place to study the conversion of gluons into the light hadron.
Because there is only one light hadron in the final state, there is no
complication due to complicated interactions between light hadrons, and
it is relatively easy to determine the conversion of gluons
into the light hadron. In this work, we propose to study the radiative decay
into $f_2(1270)$, which is a spin-2 particle and has the mass $m=1.27$GeV.
\par
The radiative decay of $J/\Psi$ was observed long time ago, and an analysis
of the polarization of $f_2$ was also performed\cite{EJ}, where the main decay mode
$f_2\to \pi+\pi$ was used. Recently the decay of $\Upsilon$ was also observed by
CLEO\cite{CU}, the branching ratio was determined.
\par
In the decay the initial state is a heavy quarkonium, say $\Upsilon$,
which can be taken as a bound state of a $b$- and $\bar b$-quark
as an approximation. Then
the radiative decay of $\Upsilon$ undergoes as the following:
the quarkonium will be annihilated into a real photon and gluons, the gluons
will subsequently be converted into the tensor meson $f_2$. In the heavy quark
limit, the $b$-and $\bar b$-quark moves with a small velocity $v$, hence
an expansion in $v$ can be employed, nonrelativistice QCD(NRQCD) can be used
to describe the nonperturbative effect related to $\Upsilon$\cite{BBL}. Also in the limit,
the tensor meson $f_2$ has a large momentum, this enables an expansion in twist
to characterize the gluonic conversion into $f_2$, the conversion is then described
by a set of distribution amplitudes of gluons. The large
momentum of $f_2$ requires that the gluons should be hard, hence
the emission of the gluons can be handled by perturbative theory.
The above discussion implies that
we may factorize the decay amplitude into three parts: the first part consists
of matrix elements of NRQCD representing the nonperturbative effect related
to $\Upsilon$, the second part consists of some distribution amplitudes, which
are for the gluonic conversion into $f_2$, the third part consists of some
coefficients, which can be calculated with perturbative theory for
the $b\bar b$-pair annihilated into gluons and a real photon. This work is
an attempt to obtain a factorized form for the decay amplitude at tree level,
in which
we analyse the gluon conversion up to twist-3 level and give the definition
of the corresponding distribution amplitudes. At loop levels, if the factorization
still holds, the perturbative coefficients should be free from infrared singularities,
and all nonperturbative effects should be contained in the NRQCD matrix elements
and in the distribution amplitudes. To warranty this one needs to prove the
factorization at loop levels, which is beyond the current work.
\par
The radiative decay was studied in the framework of a nonrelativistic
quark model\cite{KK}, in which, not only a nonrelativistic description is employed
for the initial quarkonium, but also for the tensor meson. The later
is not well justified. Instead of this we use a set of gluonic distribution
amplitudes for $f_2$, which are relativistic, gauge invariant, and universal.
Hence once information of them is extracted from one experiment, it can also be
used for predictions for other experiments. For example,
recently hard exclusive production of $f_2$ was studied\cite{BK},
in which the same distribution amplitudes appeared.
In this work we will use
our results for the two decays: $\Upsilon\to \gamma +f_2$ and
$J/\Psi\to \gamma +f_2$, and an agreement between experiment and our results
can be reached under certain conditions.
\par
Part of our results may also used for the radiative decay into $\eta$. It turns
out that the decay amplitude is at twist-4 level. Without a complete analysis
we can still predict the decay width for $\Upsilon\to\gamma+\eta$.
We will briefly discuss this
decay mode.
\par
Our work is organized as the following: In Sect.2 we introduce our notation
and perform the analysis at twist-2 level. In sect. 3 we perform the analysis
at twist-3 level. In Sect. 4 we make a comparison
of our results with experiment and give a brief discussion for the decay into $\eta$.
Sect.5 is our conclusion.
\par
Throughout of our work we take nonrelativistic normalization for the quarkonium
state and for heavy quarks.

\vskip 20pt \noindent
{\bf 2. Notations and results at order of twist-2}

\vskip20pt
We consider the exclusive decay of $\Upsilon$ in its rest-frame:
\begin{equation}
 \Upsilon (P)\rightarrow \gamma (q) +f_2(k),
\end{equation}
where the momenta are given in the brackets. Since $f_2$ is a spin-2
particle, its polarization is described a symmetric, trace-less tensor
$\varepsilon^{\mu\nu}(k,\lambda)$, where $\lambda$ is
the helicity of $f_2$.  This tensor can be constructed by introducing
polarization vectors $\omega$'s. We take a coordinate system in which
$f_2$ moves in the direction of the $z$-axis. In this system the vectors
take the form
\begin{equation}
\omega^\mu (1)=\frac{-1}{\sqrt 2}(0,1,i,0),\ \ \
\omega^\mu (-1)=\frac{1}{\sqrt 2}(0,1,-i,0),\ \ \
\omega^\mu (0) =\frac{1}{m} (\vert {\bf k}\vert,0,0, k^0).
\end{equation}
With these vectors and with Clebsch-Gorden coefficients one can construct
$\varepsilon^{\mu\nu}(k,\lambda)$ for $\lambda =0, \pm 1 \pm 2$.
At the leading order
of QED the $S$-matrix element for the decay is
\begin{equation}
\langle \gamma f_2(\lambda) |S|\Upsilon\rangle =
 -ie Q_b\varepsilon^{*\mu} \cdot\int d^4ze^{iq\cdot
z}\langle f_2(\lambda)|\bar{b}(z)\gamma _\mu b(z)\vert \Upsilon \rangle ,
\end{equation}
where $Q_b$ is the electric charge of $b$-quark in unit of $e$, $b(x)$ is
the Dirac field for $b$-quark, $\varepsilon^{*\mu}$ is the polarization vector
of the photon and $\lambda$ is the helicity of $f_2$.
In Eq.(3) we only take the part of $b$-quark in the electric current into
account. If two gluons are emitted by the $b$- or $\bar{b}$-quark,
we obtain the corresponding contribution to the $S$-matrix element
\begin{eqnarray}
 w_2 &=&-i\frac 12e Q_bg_s^2 \varepsilon^{*\mu} \int
d^4xd^4yd^4ze^{iq\cdot z}  \nonumber \\
&&\langle f_2(\lambda)|T\left[ \bar{b}(x)\gamma \cdot G(x)b(x)\bar{b}(y)\gamma \cdot
G(y)b(y)\bar{b}(z)\gamma ^\mu b(z)\right] |\Upsilon \rangle ,
\end{eqnarray}
where $G(x)$ is the gluon field. Using Wick-theorem we can calculate the $T$
-ordered product and we only keep those terms in which one b-field and one $
\bar{b}$-field remain uncontracted. Then the matrix element takes a
complicated form and can be written in a short notation:
\begin{eqnarray}
 w_2 &=&-i\frac 12e Q_bg_s^2 \varepsilon^*_\rho \int
d^4xd^4yd^4zd^4x_1d^4y_1 e^{iq\cdot z}\langle f_2(\lambda) |G_\mu ^a(x)G_\nu
^b(y)|0\rangle  \nonumber \\
&&\langle 0|\bar{b}_j(x_1)b_i(y_1)|\Upsilon\rangle \cdot M_{ji}^{\mu \nu \rho
,ab}(x,y,x_1,y_1,z),
\end{eqnarray}
where $M_{ji}^{\mu \nu \rho ,ab}(x,y,x_1,y_1,z)$ is a known function, $i$
and $j$ stand for Dirac- and color indices, $a$ and $b$ is the color of
gluon field. The above equation can be generalized to emission of arbitrary
number $n$ of gluons, the corresponding contribution is $w_n$,
then the $S$-matrix element is the sum
\begin{equation}
\langle \gamma f_2(\lambda)|S|\Upsilon \rangle =\sum_n w_n.
\end{equation}
In each contribution there
is the same matrix element $\langle 0|\bar{b}_j(x)b_i(y)|\Upsilon\rangle $.
For this matrix element the expansion in $v$ can be now performed, the
result is:
\begin{equation}
\langle 0|\bar{b}_j(x)b_i(y)|\Upsilon \rangle =-\frac 16(P_{+}\gamma ^\ell
P_{-})_{ij}\langle 0|\chi ^{\dagger }\sigma ^\ell \psi |\Upsilon \rangle
e^{-ip\cdot (x+y)}+{\cal O}(v^2),
\end{equation}
where $\chi ^{\dagger }(\psi )$ is the NRQCD field for $\bar{b}(b)$ quark
and
\begin{eqnarray}
P_{\pm } &=&(1\pm \gamma ^0)/2,  \nonumber \\
p^\mu &=&(m_b,0,0,0),
\end{eqnarray}
where $m_b$ is the pole-mass of the $b$-quark.
The leading order of the matrix element is ${\cal O}(v^0)$, we will
neglect the contribution from higher orders and the momentum of $\Upsilon $ is
then approximated by $2p$. It should be noted that effects at higher order
of $v$ can be added with the expansion in Eq.(7). Taking the result in
Eq.(7) we can write the $S$-matrix element as:
\begin{eqnarray}
 w_2 &=&i\frac 1{24}e Q_bg_s^2(2\pi )^4\delta
^4(2p-k-q)\varepsilon^*_\rho
\langle 0|\chi ^{\dagger }\sigma ^\ell \psi |\Upsilon  \rangle \nonumber \\
&&\int\frac{d^4q_1}{(2\pi )^4} \Gamma_{2\mu\nu}(k,q_1,\lambda)
\cdot R_2^{\mu \nu \rho \ell }(p,k,q_1),
\end{eqnarray}
with
\begin{equation}
\Gamma_2^{\mu\nu}(k,q_1,\lambda) =\int dx^4 e^{-iq_1\cdot x}
 \langle f_2(\lambda) \vert G^{a,\mu}(x) G^{a,\nu}(0) \vert 0\rangle.
\end{equation}
To obtain Eq.(9) we have used the color-symmetry and the translational
covariance. $\Gamma_2^{\mu\nu}(k,q_1)$ contains all nonperturbative effect
related to $f_2$, while $R_2^{\mu \nu \rho \ell }(p,k,q_1)$ is a perturbative
function, whose physical interpretation is that it is the amplitude for a $^3S_1$
$b\bar{b}$ pair emitting two gluons and a real photon, and the quarks
have the same momentum $p$. The contribution of $w_2$ may be
represented by the diagrams given in Fig. 1. In Fig.2 we give the diagrams
corresponding to emission of three gluons, their contributions $w_3$ will be analysed
in the next section. Emissions of more than three gluons will lead to contributions
which are at orders higher than those of twist-3, and will not be considered
in this work.
\par
To perform the twist expansion it is convenient to introduce the light-cone
coordinate system, in which a vector $A$ is given by $A^\nu =(A^+, A^-, A^1,A^2)
=(A^+,A^-,A_T)$, the component $A^+$ and $A^-$ is related to $A^0$ and
$A^3$ in the usual coordinate system by
\begin{equation}
 A^+ =\frac{1}{\sqrt 2} (A^0+A^3),\ \ \  A^- =\frac{1}{\sqrt 2} (A^0-A^3).
\end{equation}
In the light-cone coordinate system the momentum $k$ of $f_2$
takes the form:
\begin{equation}
 k^\mu = (k^+, k^-, 0,0), \ \ \ \ k^-=\frac{m^2}{2k^+}.
\end{equation}
In the heavy quark limit, $k^+$ is very large, while $k^-$ goes to zero.
We also define two vectors $n$ and $l$ and a tensor $d_T$:
\begin{eqnarray}
 n^\mu  &=& (0,1,0,0),\ \ \ \ l^\mu =(1,0,0,0),\nonumber \\
 d_T^{\mu\nu}&=& g^{\mu\nu}-n^\mu l^\nu-n^\nu l^\mu.
\end{eqnarray}
Different components of a vector can be projected out with these vectors and the
tensor:
\begin{equation}
 A^+ =n\cdot A,\ \ \ A^-=l\cdot A,\ \ \  A_T^\mu = d_T^{\mu\nu}A_\nu.
\end{equation}
The light-cone gauge is chosen in our work and is defined by
\begin{equation}
 n\cdot G(x) =G^+(x) =0.
\end{equation}
In this gauge a set of components of the gluon field strength takes a
simple form:
\begin{equation}
G^{+\mu}(x)=\partial^+ G^\mu (x) =\frac{\partial }{\partial x^-} G^\mu (x).
\end{equation}
\par
The nonperturbative object $\Gamma_2^{\mu\nu}$ characterize the conversion
of two gluons into $f_2$. The $x$-dependence of the matrix element
$\langle f_2(k,\lambda) \vert G^{a,\mu}(x) G^{a,\nu}(0) \vert 0\rangle$
is controlled by different scales: the $x^-$-dependence is controlled
by $k^+$, while the $x^+$- and $x_T$-dependence is controlled by
the scale $\Lambda_{QCD}$ or $k^-$, which are small in comparison with $k^+$.
Because of these small scales we can expand the matrix element in $x^+$ and in
$x_T$. With this expansion we obtain
\begin{eqnarray}
\Gamma_2^{\mu\nu}(k,q_1,\lambda) &=& (2\pi)^3 \delta(q_1^-) \delta^2 (q_{1T})
\cdot \int dx^- e^{-iq_1^+x^-}
\langle f_2(\lambda) \vert G^{a,\mu}(x^-) G^{a,\nu}(0) \vert 0\rangle
\nonumber\\
&& +i(2\pi)^3 \delta(q_1^-) \frac{\partial}{\partial q^\rho_{1T}}\delta^2 (q_{1T})
\cdot \int dx^- e^{-iq_1^+x^-}
\langle f_2(\lambda) \vert \partial_T^\rho G^{a,\mu}(x^-) G^{a,\nu}(0) \vert 0\rangle
\nonumber\\
 && +i(2\pi)^3 \frac{\partial}{\partial q_1^-}\delta(q_1^-) \delta^2 (q_{1T})
\cdot \int dx^- e^{-iq_1^+x^-}
\langle f_2(\lambda) \vert \partial^-G^{a,\mu}(x^-) G^{a,\nu}(0) \vert 0\rangle
\nonumber\\
&& +\cdots\cdots,
\end{eqnarray}
where we introduced a short notation:
\begin{equation}
 G^\mu (x^-) =G^\mu(x^-n).
\end{equation}
In Eq.(17), the leading twist of the first term is 2, while
the leading twist of the second term is 3. By considering Lorentz
covariance and varying the vector $n$\cite{EFP} one can show that the derivative
$\partial^-$ in the third term is related to $\partial^2_T$, hence the leading
twist of the third term is 4. The $\cdots\cdots$ stand for contributions
which have twist more than 3 and will be neglected. It should be noted that the
expansion in Eq.(17) is equivalent to the collinear expansion around $q_1^\mu
=(q_1^+,0,0,0)$ for $R_2^{\mu \nu \rho \ell }(p,k,q_1)$.
\par
The leading-twist contribution in the first term is specified by that
the indices $\mu$ and $\nu$ are all transversal indices, i.e., $\mu$
or $\nu=1,2$. By considering the covariance under Lorentz boost
along the $z$-axis and the invariance of
rotations around the $z$-axis we find at twist-2 for $\lambda=0$:
\begin{equation}
\Gamma_2^{\mu\nu} (k,q_1,0) =  (2\pi)^4 \delta(q_1^-) \delta^2
(q_{1T}) \cdot \frac{1}{2k^+x_1(x_1-1)} d_T^{\mu\nu} F_0(x_1),
\end{equation}
where $q_1^+=x_1k^+$, $F_0(x_1)$ is the gluonic distribution
amplitude for $\lambda =0$ and is given by
\begin{equation}
F_0(x_1)= \frac{1}{2\pi k^+} \int dx^-e^{-ix_1k^+x^-}
\langle f_2(0) \vert G^{a,+\mu}(x^-) G^{a,+}_{\ \ \ \mu} (0) \vert 0\rangle
\end{equation}
It should be noted that
\begin{equation}
 \frac {1}{\sqrt 6} d_T^{\mu\nu} =\varepsilon_T^{*\mu\nu}(0)
 =\varepsilon^*_{\mu'\nu'}(0)d_T^{\mu'\mu}d_T^{\nu'\nu}.
\end{equation}
\par
At twist-2 level $\Gamma^{\mu\nu} (k,q_1,\lambda)$ can also be
nonzero for $\lambda =\pm 2$. The contribution for $\lambda=2$
 can be written as
\begin{equation}
\Gamma_2^{\mu\nu} (k,q_1,2) =  (2\pi)^4 \delta(q_1^-) \delta^2 (q_{1T})
\cdot \frac{1}{k^+x_1(x_1-1)} \varepsilon^{*\mu\nu}(2)  F_2(x_1),
\end{equation}
where $F_2(x_1)$ is the gluonic distribution
amplitude for $\lambda =2$ and is defined by:
\begin{equation}
F_2(x_1)= \frac{1}{2\pi k^+} \int dx^-e^{-ix_1k^+x^-}
\langle f_2(2) \vert G^{a,+\mu}(x^-) G^{a,+\nu} (0) \vert 0\rangle
 \omega_\mu (1) \omega_\nu (1)
\end{equation}
The polarization vector $\varepsilon^{\mu\nu}(2)$ is given by
\begin{equation}
\varepsilon^{\mu\nu}(2) =\omega^\mu (1) \omega^\nu (1).
\end{equation}
\par
At twist-2 level $f_2$ can not have the helicity $\lambda=1$, because the
two gluons are collinear with a relative angular momentum which is zero
and they have Bose-symmetry.
With these results for $\Gamma^{\mu\nu} (k,q_1,\lambda )$ and with the explicit
form of $R_2^{\mu \nu \rho \ell }(p,k,q_1)$ we obtain the $S$-matrix element with
$\lambda =0$ and with $\lambda =2$:
\begin{eqnarray}
\langle \gamma f_2(0) |S|\Upsilon\rangle &=&
  \frac i{6}e Q_bg_s^2(2\pi )^4\delta^4(2p-k-q)\varepsilon^{*\ell}
\langle 0|\chi ^{\dagger }\sigma ^\ell \psi |\Upsilon  \rangle \
 \cdot \frac {1}{m_b^2} \cdot T_0,
\nonumber\\
\langle \gamma f_2(2) |S|\Upsilon\rangle &=&
  \frac i{6}e Q_bg_s^2(2\pi )^4\delta^4(2p-k-q)(-\varepsilon^*_\mu\omega^{*\mu}(1))
\omega^{*\ell}(1) \langle 0|\chi ^{\dagger }\sigma ^\ell \psi |\Upsilon  \rangle \
 \cdot \frac {1}{m_b^2} \cdot T_2,
\end{eqnarray}
with
\begin{eqnarray}
T_0 &=& \int dx_1 \frac{ 8m_b^2(x_1-1) +m^2(1-2x_1)}
      { 2x_1(1-x_1) (4m_b^2 (x_1-1)+m^2(1-2x_1))}\cdot F_0(x_1)
\nonumber\\
T_2 &=& \int dx_1 \frac {m^2}{x_1(1-x_1)(4m_b^2(x_1-1)+m^2(1-2x_1))}
   \cdot F_2(x_1).
\end{eqnarray}
In the calculation we have kept the mass of $f_2$. The effect of the mass
is usually regarded as a correction to the leading twist effect. From the above
results we can see that the amplitude with $\lambda =2$ is suppressed
by $(m/m_b)^2$, although it is a leading twist contribution. At twist-2
two gluonic distribution amplitudes characterize the gluon conversion into
$f_2$ with the helicity $\lambda=0$ and $\lambda=2$, respectively. They
are defined in the light-cone gauge and are invariant under gauge transformations,
which respect the gauge condition in Eq.(15). In other gauges a gauge link between
the two gluonic operators in $F_0$ and in $F_2$ should be added to make
them gauge invariant.  Because of the momentum-conservation in the $+$-direction,
the functions $F_0(x)$ and $F_2(x)$ becomes zero if $x>1$ or $x<0$.
These functions are defined at certain renormalization scale $\mu$, hence they depends not only
on $x$ but also on $\mu$. The evolution of these
twist-2 operators are well known.
Our results given above receive corrections from
orders of higher twist and the corrections are suppressed by $(\Lambda/m_b)^2$
where $\Lambda$ can be $\Lambda_{QCD}$ or $k^-$.
\par
With these results in Eq.(25) and Eq.(26) we complete the analysis at twits-2
level. The contributions at twist-3 come from the first and the second term in
Eq.(17) and also from contributions with emission of three gluons, which
are represented by Fig.2. We will analyse these contributions in the next section
and show that the amplitude at twist-3 level is nonzero for $\lambda=1$.
\par\vskip 20pt
\noindent
{\bf 3. Twist-3 Contributions}
\par\vskip20pt
In this section we calculate the contribution at order of twist-3. It turns out
that at this order the decay amplitude is nonzero only for $\lambda=1$. We will
neglect in this section the effect of $m$. The effect may be included and
including it will results in complicated expressions. We start
with Eq.(17). There are two terms which are twist-3 contributions,
one term is specified by that
one of the indices $\mu$ and $\nu$ is $-$ in the first term, while another
term is specified by the indices $\mu$ and $\nu$ to be all transversal in the second
term.  Examining their Lorentz structure we find it is nonzero for $\lambda=\pm 1$.
Hence we obtain the leading-twist contribution for $\lambda =1$ as
\begin{eqnarray}
\Gamma_2^{\mu\nu}(k,q_1,1) &=&\Gamma_2^{(1)\mu\nu}(k,q_1)
                           +\Gamma_2^{(2)\mu\nu}(k,q_1)
\nonumber\\
\Gamma_2^{(1)\mu\nu}(k,q_1)&=&
(2\pi)^3 \delta(q_1^-) \delta^2 (q_{1T})
\nonumber\\
&\cdot& \int dx^- e^{-iq_1^+x^-}
\langle f_2(1) \vert n^\mu G^{a,-}(x^-) G_T^{a,\nu}(0)
             +n^\nu G_T^{a,\mu}(x^-) G^{a,-}(0)\vert 0\rangle
\nonumber\\
\Gamma_2^{(2)\mu\nu}(k,q_1)&=&
i(2\pi)^3 \delta(q_1^-) \frac{\partial}{\partial q^\rho_{1T}}\delta^2 (q_{1T})
\cdot \int dx^- e^{-iq_1^+x^-}
\langle f_2(1) \vert \partial_T^\rho G_T^{a,\mu}(x^-) G_T^{a,\nu}(0) \vert 0\rangle,
\end{eqnarray}
where $G_T^{a,\mu}=d_T^{\mu\nu}G^a_\nu$. For $\Gamma_2^{(1)\mu\nu}(k,q_1)$ we can define
a distribution amplitude
\begin{equation}
G_0 (x) = \frac {1}{2\pi} \int dx^- e^{-ixk^+x^-}
\langle f_2(\lambda) \vert G_T^{a,+-}(x^-) G_T^{a,\mu}(0)\vert 0\rangle\omega_\mu(1).
\end{equation}
It should be noted that $G^{-}$ in the light cone gauge is not an independent
dynamical freedom. Its relation to the gauge field $G_T^{\mu}$ and quark fields
can be obtained by solving equation of motion of QCD. One can obtain
that $G^{+-}$ can be expressed by $D_T^\mu G_{T\mu}$ and the color-charge current
of quarks, where $D^\mu$ is the covariant derivative in the adjoint
representation of $SU(3)$:
\begin{equation}
\left [ D^\mu \right ]_{ab} =\partial^\mu\delta_{ab} +g_s f_{abc} G^{c,\mu},
\ \ {\rm for\ } a,b=1, \cdots, 8.
\end{equation}
Because of $D_T^\mu G_{T\mu}$ the contribution is a twist-3 contribution. We
will not try to give the result for
$G^-$, but leave it with the compact form.
Expressing
$\Gamma_2^{(1)\mu\nu}(k,q_1)$ with $G_0(x)$ we obtain the contribution to the
$S$-matrix element $\langle \gamma f_2(1) |S|\Upsilon\rangle_0$:
\begin{eqnarray}
\Gamma_2^{(1)\mu\nu}(k,q_1)&=& \frac{2\pi}{k^+(k-q_1)^+}
        \{ n^\mu \omega^{*\nu} G_0(x_1)+n^\nu\omega^{*\mu} G_0(1-x_1)\},
\nonumber\\
 \langle \gamma f_2(1) |S|\Upsilon\rangle_0&=&
\frac i{6}e Q_bg_s^2(2\pi )^4\delta^4(2p-k-q)\varepsilon^*\cdot\omega^{*}(1)
 l^i\langle 0|\chi ^{\dagger }\sigma ^i \psi |\Upsilon  \rangle
 \frac{1}{{\sqrt 2}m_b^3}
\nonumber\\
&&\cdot\int dx \frac{2}{x(1-x)} G_0(x).
\end{eqnarray}
\par
The Lorentz structure of the integral in $\Gamma_2^{(2)\mu\nu}(k,q_1)$ can be
decomposed as:
\begin{eqnarray}
C^{\mu\nu\rho} &=& \int dx^- e^{-iq_1^+x^-}
\langle f_2(1) \vert \partial_T^\rho G_T^{a,\mu}(x^-) G_T^{a,\nu}(0) \vert 0\rangle
\nonumber\\
&=& \frac{1}{k^+(q_1-k)^+}\int dx^- e^{-iq_1^+x^-}
\langle f_2(1) \vert \partial_T^\rho
 G_T^{a,+\mu}(x^-) G_T^{a,+\nu}(0) \vert 0\rangle
\nonumber\\
&=& \frac{k^+}{4k^+(q_1-k)^+}\Big\{ ( \omega^{*\mu}d_T^{\rho\nu} +\omega^{*\nu}d_T^{\rho\mu}
    -3\omega^{*\rho}d_T^{\mu\nu}) K_0(x_1)
\nonumber\\
&+& \frac{1}{4} ( \omega^{*\rho}d_T^{\mu\nu} +\omega^{*\nu}d_T^{\rho\mu}
    -3\omega^{*\mu}d_T^{\rho\nu}) K_1(x_1)
\nonumber\\
&+&\frac{1}{4} ( \omega^{*\rho}d_T^{\mu\nu} +\omega^{*\mu}d_T^{\rho\nu}
    -3\omega^{*\nu}d_T^{\rho\mu}) K_2(x_1)\Big\},
\end{eqnarray}
where the functions are given by
\begin{eqnarray}
K_0(x) &=& \frac{1}{k^+} \int dx^- e^{-ixk^+x^-}
  \langle f_2(1) \vert \partial_T^\rho
 G_T^{a,+\mu}(x^-) G^{a,+}_{T\ \ \mu}(0) \vert 0\rangle\omega_\rho(1),
\nonumber\\
K_1(x) &=& \frac{1}{k^+} \int dx^- e^{-ixk^+x^-}
  \langle f_2(1) \vert \partial_T^\mu
 G_T^{a,+\rho}(x^-) G^{a,+}_{T\ \ \mu}(0) \vert 0\rangle\omega_\rho(1),
\nonumber\\
K_2(x) &=& \frac{1}{k^+} \int dx^- e^{-ixk^+x^-}
  \langle f_2(1) \vert \partial_T^\mu
 G^{a,+}_{T\ \ \mu}(x^-) G_T^{a,+\rho}(0) \vert 0\rangle\omega_\rho(1),
\end{eqnarray}
and they have the following property:
\begin{equation}
K_0(x)=-K_0(1-x),\ \ \ K_1(x)=-K_2(1-x).
\end{equation}
With these results we obtain the contribution to the $S$-matrix element
from $\Gamma_2^{(2)\mu\nu}(k,q_1)$:
\begin{eqnarray}
\langle \gamma f_2(1) |S|\Upsilon\rangle_1&=&
\frac i{6}e Q_bg_s^2(2\pi )^4\delta^4(2p-k-q)\varepsilon^*\cdot\omega^{*}(1)
 l^i\langle 0|\chi ^{\dagger }\sigma ^i \psi |\Upsilon  \rangle
\nonumber\\
 && \frac{1}{{\sqrt 2}m_b^3}\frac{1}{2\pi} \int dx_1 \frac{1-2x_1}{(1-x_1)^2x_1^2}
K_0(x_1),
\nonumber\\
\langle \gamma f_2(1) |S|\Upsilon\rangle_2&=&
\frac i{6}e Q_bg_s^2(2\pi )^4\delta^4(2p-k-q)\varepsilon^*\cdot\omega^{*}(1)
 l^i\langle 0|\chi ^{\dagger }\sigma ^i \psi |\Upsilon  \rangle
 \nonumber\\
&& \frac{1}{{\sqrt 2}m_b^3}\frac{1}{2\pi}
\int dx_1 \frac {2}{x_1(1-x_1)^2} K_2(x_1).
\end{eqnarray}
In the above results $G_0(x)$ is gauge invariant, hence the contribution
$\langle \gamma f_2(1) |S|\Upsilon\rangle_0$ also does. But the function
$K_0(x)$ and $K_1(x)$ is not gauge invariant, the contributions
in Eq.(34) also are not gauge invariant. The gauge invariance can be achieved
by considering the emission of three gluons, which we will consider in
the below.
\par
The contribution from type of diagrams given in Fig.2a can be written
after the expansion in Eq.(7):
\begin{eqnarray}
 w_{3a} &=&\frac i{6}e Q_bg_s^3(2\pi )^4\delta
^4(2p-k-q)\varepsilon^*_\rho
\langle 0|\chi ^{\dagger }\sigma ^\ell \psi |\Upsilon  \rangle
\cdot \frac{i}{4}\cdot \frac {1}{3!}\cdot
\nonumber  \\
&&\int\frac{d^4q_1}{(2\pi )^4}\frac {dq_2^4}{(2\pi)^4}
 \Gamma_{3a\mu_1\mu_2\mu_3}(k,q_1,q_2,\lambda)
\cdot R_{3a}^{\mu_1\mu_2\mu_3\rho\ell}(p,k,q_1,q_2),
\end{eqnarray}
with
\begin{equation}
\Gamma_{3a}^{\mu_1\mu_2\mu_3}(k,q_1,q_2,\lambda)
 = \int dx^4dy^4 e^{-iq_1\cdot x}e^{-iq_2\cdot y} f_{abc}
 \langle f_2(\lambda) \vert G^{a,\mu_1}(x)G^{b,\mu_2}(y) G^{c,\mu_3}(0) \vert 0\rangle.
\end{equation}
In Eq.(35) the factor $i/4$ comes from the trace over color indices, the three gluons
forms a C-even state, hence the contribution is proportional to $f_{abc}$. The factor
$1/3!$ is a statistical factor. The leading order of $w_{3a}$ is twist-3, its
contribution is given by keeping the $x^-$- and $y^-$dependence in the matrix element
only and by all gluon fields being transversal. By examining the Lorentz structure
of this contribution it is nonzero only for $\lambda=\pm 1$. The leading term
of $\Gamma_{3a}$ for $\lambda =1$ can be written:
\begin{equation}
\Gamma_{3a}^{\mu_1\mu_2\mu_3}(k,q_1,q_2,1) =
(2\pi)^6 \delta(q_1^-)\delta(q_2^-) \delta^2(q_{1T})\delta^2(q_{2T})
C_{3a}^{\mu_1\mu_2\mu_3},
\end{equation}
where the tensor $C_{3a}^{\mu_1\mu_2\mu_3}$ is given by
\begin{eqnarray}
C_{3a}^{\mu_1\mu_2\mu_3} &=& \frac{1}{4}
( \omega^{*\mu_2}d_T^{\mu_1\mu_3} +\omega^{*\mu_3}d_T^{\mu_1\mu_2}
    -3\omega^{*\mu_1}d_T^{\mu_2\mu_3}) H_0(x_1,x_2)
\nonumber\\
&+& \frac{1}{4} ( \omega^{*\mu_1}d_T^{\mu_2\mu_3} +\omega^{*\mu_3}d_T^{\mu_1\mu_2}
    -3\omega^{*\mu_2}d_T^{\mu_1\mu_3}) H_1(x_1,x_2)
\nonumber\\
&+&\frac{1}{4} ( \omega^{*\mu_1}d_T^{\mu_2\mu_3} +\omega^{*\mu_2}d_T^{\mu_1\mu_3}
    -3\omega^{*\mu_3}d_T^{\mu_1\mu_2}) H_2(x_1,x_2).
\end{eqnarray}
The functions are defined by:
\begin{eqnarray}
H_0(x_1,x_2) &=& \int dx^-dy^- e^{-ix_1k^+x^--ix_2k^+y^-} f_{abc}
\langle f_2(1) \vert G_T^{a,\mu}(x^-)G^{b,\nu}(y^-) G^{c}_\nu(0) \vert 0\rangle
\omega_\mu (1),
\nonumber\\
H_1(x_1,x_2) &=& \int dx^-dy^- e^{-ix_1k^+x^--ix_2k^+y^-} f_{abc}
\langle f_2(1) \vert G^{a,\nu}(x^-)G_T^{b,\mu}(y^-) G^{c}_\nu(0) \vert 0\rangle
\omega_\mu (1),
\nonumber\\
H_3(x_1,x_2) &=& \int dx^-dy^- e^{-ix_1k^+x^--ix_2k^+y^-} f_{abc}
\langle f_2(1) \vert G^{a,\nu}(x^-)G^b_\nu(y^-) G_T^{c,\mu}(0) \vert 0\rangle
\omega_\mu (1),
\end{eqnarray}
where the variable $x_1$ and $x_2$ are related to $q_1^+$ and $q_2^+$ by
\begin{equation}
   q_1^+=x_1k^+\ {\rm and}\ \ q_2^+=x_2k^+,
\end{equation}
respectively. The functions are related to each other:
\begin{equation}
H_0(x_1,x_2) =-H_1(x_2,x_1),\ \ H_2(x_1,x_2)=H_0(1-x_1-x_2,x_1),
\end{equation}
and there are also some identities among them, one of them, which will be used
later, is given by
\begin{equation}
H_2(x_1,x_2) =-H_2(x_2,x_1).
\end{equation}
After a tedious calculation we obtain the leading-twist contribution of
$\Gamma_{3a}$ to the $S$-matrix element
\begin{eqnarray}
\langle \gamma f_2(1) |S|\Upsilon\rangle_3&=&
\frac i{6}e Q_bg_s^2(2\pi )^4\delta^4(2p-k-q)\varepsilon^*\cdot\omega^{*}(1)
 l^i\langle 0|\chi ^{\dagger }\sigma ^i \psi |\Upsilon  \rangle
 \nonumber\\
&& \frac{ig_s}{{\sqrt 2}m_b^3}\frac{(k^+)^2}{(2\pi)^2}
\int dx_1 dx_2\frac{x_1+2x_2-1}{x_1x_2-x_1+x_2^2-x_2}\cdot H_0(x_1,x_2).
\end{eqnarray}
We find that this contribution can be combined with
$\langle \gamma f_2(1) |S|\Upsilon\rangle_1$ to form a gauge-invariant contribution.
For doing this, we insert an identity into
$\langle \gamma f_2(1) |S|\Upsilon\rangle_1$:
\begin{equation}
 \frac{k^+}{2\pi} \int dx_2 \int dy^- e^{-ix_2k^+y^-}
  =\int dx_2 \delta(x_2)=1
\end{equation}
we obtain
\begin{eqnarray}
\langle \gamma f_2(1) |S|\Upsilon\rangle_1 &=& (\cdots)
 \frac{1}{(2\pi)^2} \int dx_1dx_2
  \frac{1-2x_1}{(1-x_1)^2x_1^2}
\nonumber\\
 && \int dx^-dy^-
   e^{-ix_1k^+x^--ix_2k^+y^-}
\langle f_2(1) \vert \partial_T^\rho
 G_T^{a,+\mu}(x^-) G^{a,+}_{T\ \ \mu}(0) \vert 0\rangle\omega_\rho(1),
\end{eqnarray}
where $(\cdots)$ stands for some common factors. For
$\langle \gamma f_2(1) |S|\Upsilon\rangle_3$ we make the variable change
$x_1\to x_2$ and $x_2\to x_1$ and write some gluon fields into field
strength tensors:
\begin{eqnarray}
\langle \gamma f_2(1) |S|\Upsilon\rangle_3 &=& (\cdots)
\frac{-g_s}{(2\pi)^2} \int dx_1 dx_2 \frac{x_2+2x_1-1}{x_1(1-x_1-x_2)
 (x_1x_2-x_2+x_1^2-x_1)}
\nonumber\\
&& \int dx^-dy^- e^{-ix_1k^+x^--ix_2k^+y^-} f_{abc}
\langle f_2(1) \vert G_T^{a,\mu}(y^-)G^{b,+\nu}(x^-) G^{c,+}_{\ \ \nu}(0) \vert 0\rangle
\omega_\mu (1),
\end{eqnarray}
where $(\cdots)$ stands for the same common factors as in Eq.(45).
We note that by setting $x_2=0$ the function in the first line of
Eq.(46) is the same in the first line of Eq.(45). Because of the identity
in Eq.(44) we can combine the two terms together as
\begin{eqnarray}
\langle \gamma f_2(1) |S|\Upsilon\rangle_1&+& \langle \gamma f_2(1) |S|\Upsilon\rangle_3
\nonumber\\
&=& (\cdots) \frac{1}{(2\pi)^2}\int dx_1dx_2\frac{x_2+2x_1-1}{x_1(1-x_1-x_2)
 (x_1x_2-x_2+x_1^2-x_1)}
\nonumber\\
&& \int dx^-dy^- e^{-ix_1k^+x^--ix_2k^+y^-}
\langle f_2(1) \vert \partial_T^\mu
 G^{a,+\nu}(x^-) G^{a,+}_{\ \ \ \nu}(0)
\nonumber\\
&& -g_s f_{abc}G_T^{a,\mu}(y^-)G^{b,+\nu}(x^-) G^{c,+}_{\ \ \ \nu}(0) \vert 0\rangle
\omega_\mu (1).
\end{eqnarray}
We note that the two terms in the matrix element of the above equation can be
written as a one term with the covariant derivative $D_T^\mu(y)$ which
is defined as
\begin{equation}
\left [ D^\mu (y)\right ]_{ab} =\partial^\mu\delta_{ab} +g_s f_{abc} G^{c,\mu}(y),
\ \ {\rm for\ } a,b=1, \cdots, 8.
\end{equation}
With this observation we can define our second distribution amplitude
$G_1(x_1,x_2)$:
\begin{equation}
G_1(x_1,x_2) = \frac{1}{(2\pi)^2} \int dx^- dy^- e^{-ix_1k^+x^--ix_2k^+y^-}
\langle f_2(1) \vert \big [D_T^\mu(y^-)
 G^{+\nu}(x^-)\big ]^a  G^{a,+}_{\ \ \ \nu}(0) \vert 0\rangle
\omega_\mu (1).
\end{equation}
This amplitude is gauge invariant in the light-cone gauge. In other gauges two
gauge links must be supplied to ensure gauge invariance. With it the sum
can be written:
\begin{eqnarray}
\langle \gamma f_2(1) |S|\Upsilon\rangle_1&+& \langle \gamma f_2(1) |S|\Upsilon\rangle_3
\nonumber\\
&=& (\cdots) \frac{1}{(2\pi)^2}\int dx_1dx_2\frac{x_2+2x_1-1}{x_1(1-x_1-x_2)
 (x_1x_2-x_2+x_1^2-x_1)} \cdot G_1(x_1,x_2).
\end{eqnarray}
\par
Now we turn to the contribution represented by Fig.2b, where the three-gluon
vertex is involved. This contribution can be obtained from $w_2$ in Eq.(9)
by using perturbative theory with the three-gluon vertex for $\Gamma_2$ once. We denote the
term by using the perturbative theory for $\Gamma_2$ as $\Gamma_{23}$, which
is given by:
\begin{eqnarray}
\Gamma_{23}^{\mu\nu}(k,q_1,\lambda)  &=& ig_s f_{abc} \int dx^4 e^{-iq_1\cdot x}
\big\{  \langle f_2(\lambda) \vert \big [ iD^{\rho\mu}(q_1)
      \partial_\sigma G^a_\rho(x) G^{b,\sigma}(x) G^{c,\nu}(0)
\nonumber\\
 && -i D^{\sigma\mu}(q_1) \partial_\sigma G^a_\rho(x) G^{b,\rho}(x) G^{c,\nu}(0)
  +q_{1\sigma} D_\rho^{\ \mu}(q_1) G^{a,\sigma}(x) G^{b,\rho}(x) G^{c,\nu}(0)
 \big] \vert 0\rangle
\nonumber \\
 && + (\mu \to \nu,\ \nu\to\mu,\ q_1\to k-q_1) \big\}
\end{eqnarray}
where $D^{\mu\nu}(q_1)$ is the gluon propagator in the light-cone gauge:
\begin{equation}
D^{\mu\nu}(q) = -\frac{1}{q^2}\left ( g^{\mu\nu}
  -\frac{ n^\mu q^\nu+n^\nu q^\mu}{n\cdot q}\right ) .
\end{equation}
The leading twist of this term is 3, whose contribution comes from
the second term in Eq.(51) with $\partial_\sigma =\partial_-=\partial^+$.
Again, here the tensor meson can only have $\lambda =\pm 1$. We denote
the corresponding contribution to the $S$-matrix element as
$\langle \gamma f_2(1) |S|\Upsilon\rangle_4$, which is then given by
\begin{eqnarray}
 \langle \gamma f_2(1) |S|\Upsilon\rangle_4&=&i\frac 1{24}e Q_bg_s^2(2\pi )^4\delta
^4(2p-k-q)\varepsilon^*_\rho
\langle 0|\chi ^{\dagger }\sigma ^\ell \psi |\Upsilon  \rangle \nonumber \\
&&\int\frac{d^4q_1}{(2\pi )^4} R_2^{\mu \nu \rho \ell }(p,k,q_1)
 \big \{ (2\pi)^3\delta(q_1^-)\delta^2(q_{1T})g_s D^{-\mu}(q_1)
\nonumber\\
&& \int dx^- e^{-ik_1^+x^-} f_{abc} \langle f_2(1) \vert G^{a,+\sigma}(x^-)
   G^b_{\ \sigma}(x^-)G^{c,\nu}(0) \vert 0 \rangle
\nonumber\\
&& + (\mu \to \nu,\ \nu\to\mu,\ q_1\to k-q_1) \big\}.
\end{eqnarray}
At the first glance the above contribution may be divergent because of
the gluon propagator is singular if we set $q_1^\mu=(q_1^+,0,0,0)$. However
we find that the integrand is finite when $q_1^\mu \to(q_1^+,0,0,0)$,
and it leads to a finite
contribution which is the same if we replace the gluon propagator
by its special propagator $\hat D^{\mu\nu}(q_1)$:
\begin{equation}
 D^{\mu\nu}(q_1) \to \hat D^{\mu\nu}(q_1) = \frac{n^\mu n^\nu}{(n\cdot q_1)^2}.
\end{equation}
A discussion of the special propagator can be found in \cite{Qiu}, where it
is shown that by using the special propagator of quarks twist-4
contributions to the hadron structure functions can be conveniently
analysed.
\par
By taking care of the singularity the calculation is straightforward.
We obtain
\begin{eqnarray}
\langle \gamma f_2(1) |S|\Upsilon\rangle_4&=&
\frac i{6}e Q_bg_s^2(2\pi )^4\delta^4(2p-k-q)\varepsilon^*\cdot\omega^{*}(1)
 l^i\langle 0|\chi ^{\dagger }\sigma ^i \psi |\Upsilon  \rangle
 \nonumber\\
&& \frac{ig_s}{{\sqrt 2}m_b^3}
\int dx_1 \frac{2}{x_1^2(1-x_1)}\cdot J_0(x_1)
\end{eqnarray}
with $J_0(x_1)$ defined as
\begin{equation}
J_0(x_1) = \frac {1}{2\pi k^+} \int dx^- e^{-ix_1k^+x^-} f_{abc}
\langle f_2(1) \vert G_T^{a,+\mu}(x^-)G^{b}_{\mu}(x^-)
G^{c,+}_{\ \ \ \nu}(0) \vert 0\rangle\omega_\nu (1).
\end{equation}
\par
To ensure the gauge invariance of the $S$-matrix element, the sum
$\langle \gamma f_2(1) |S|\Upsilon\rangle_2
+\langle \gamma f_2(1) |S|\Upsilon\rangle_4$
should be gauge invariant. As the expressions for both terms stand,
it does not look like that the sum can be written in a gauge-invariant
form. A re-arrangement is needed. We first write $J_0(x_1)$ as an double
distribution amplitude by using
\begin{equation}
G^{b}_\mu (x^-) = \frac{k^+}{2\pi} \int dx_2 dy^- e^{-ix_2k^+(y-x)^-}
   G^b_\mu (y^-).
\end{equation}
Then the contribution can be written as
\begin{eqnarray}
\langle \gamma f_2(1) |S|\Upsilon\rangle_4 &=& (\cdots)
\frac{g_s}{(2\pi)^2} \int dx_1 dx_2 \frac{2}{(x_1+x_2)^2(1-x_1-x_2)}
\nonumber\\
&& \int dx^-dy^- e^{-ix_1k^+x^--ix_2k^+y^-} f_{abc}
\langle f_2(1) \vert G_T^{a,+\mu}(x^-)G^b_\mu(y^-) G^{c,+}_{\ \ \nu}(0) \vert 0\rangle
\omega_\mu (1),
\end{eqnarray}
where $(\cdots)$ stands for the same common factors as before. We add to
$\langle \gamma f_2(1) |S|\Upsilon\rangle_4$ a term
\begin{equation}
(\cdots)
\frac{g_s}{(2\pi)^2} \int dx_1 dx_2 \frac{2(1-2x_1-2x_2)}{(x_1+x_2)(1-x_1-x_2)}
  H_2(x_1,x_2),
\end{equation}
which is identically zero because of the property in Eq.(42). With this term
the contribution can be written:
\begin{eqnarray}
\langle \gamma f_2(1) |S|\Upsilon\rangle_4 &=& (\cdots)
\frac{g_s}{(2\pi)^2} \int dx_1 dx_2 f_2(x_1,x_2)
\nonumber\\
&& \int dx^-dy^- e^{-ix_1k^+x^--ix_2k^+y^-} f_{abc}
\langle f_2(1) \vert G_T^{a,+\mu}(x^-)G^b_\mu(y^-) G^{c,+}_{\ \ \nu}(0) \vert 0\rangle
\omega_\mu (1),
\nonumber\\
f_2(x_1,x_2) &=&\left\{\frac{2}{(x_1+x_2)^2(1-x_1-x_2)}
 -\frac{2(1-2x_1-2x_2)}{x_1(x_1+x_2)(1-x_1-x_2)^2} \right \}
\end{eqnarray}
By setting $x_2 =0$ the function $f_2(x_1,0)$ is the same
as that in the front of $K_2(x_1)$ in Eq. (34). With the same trick used to derive
Eq.(47) we can write the sum as
\begin{eqnarray}
\langle \gamma f_2(1) |S|\Upsilon\rangle_2&+& \langle \gamma f_2(1) |S|\Upsilon\rangle_4
\nonumber\\
&=& (\cdots) \frac{1}{(2\pi)^2}\int dx_1dx_2 f_2(x_1,x_2)
\nonumber\\
&& \int dx^-dy^- e^{-ix_1k^+x^--ix_2k^+y^-}
\langle f_2(1) \vert \partial_T^\nu
 G^{a,+}_{\ \ \ \nu}(x^-) G^{a,+\mu}(0)
\nonumber\\
&& +g_s f_{abc}G^{a,+\nu}(x^-)G^{b}_\nu(y^-) G^{c,+\mu}(0) \vert 0\rangle
\omega_\mu (1).
\end{eqnarray}
Again the two terms in the matrix element can be combined together with
the covariant derivative in Eq. (48). We can define the third distribution amplitude
$G_2(x_1,x_2)$:
\begin{equation}
G_2(x_1,x_2) = \frac{1}{(2\pi)^2} \int dx^- dy^- e^{-ix_1k^+x^--ix_2k^+y^-}
\langle f_2(1) \vert \big [D_T^\nu(y^-)
 G^{+}_{\ \nu}(x^-)\big ]^a  G^{a,+\mu}(0) \vert 0\rangle
\omega_\mu (1).
\end{equation}
The sum can be written:
\begin{equation}
\langle \gamma f_2(1) |S|\Upsilon\rangle_2+ \langle \gamma f_2(1) |S|\Upsilon\rangle_4
= (\cdots) \frac{1}{(2\pi)^2}\int dx_1dx_2 f_2(x_1,x_2) G_2(x_1,x_2)
\end{equation}
Adding different contributions together we obtain the $S$-matrix element for
$\lambda =1$ at twist-3 level:
\begin{eqnarray}
\langle \gamma f_2(1) |S|\Upsilon\rangle&=&
\frac i{6}e Q_bg_s^2(2\pi )^4\delta^4(2p-k-q)\varepsilon^*\cdot\omega^{*}(1)
 l^i\langle 0|\chi ^{\dagger }\sigma ^i \psi |\Upsilon  \rangle
 \frac{1}{m_b^2}\cdot T_1
\nonumber\\
T_1 &=& \frac{1}{\sqrt {2} m_b} \Big [
 \int dx_1 \frac{2}{x_1(1-x_1)} G_0(x_1)
 +i\int dx_1 dx_2 \big (f_1(x_1,x_2)G_1(x_1,x_2)
 \nonumber\\
 && \ \ \ +2f_2(x_1,x_2)G_2(x_1,x_2)
 \big ) \Big ],
\end{eqnarray}
where the function $f_1$ is given by
\begin{equation}
f_1(x_1,x_2) =\frac {x_2+2x_1-1}{x_1(1-x_1-x_2)(x_1x_2-x_2+x_1^2-x_1)}.
\end{equation}
\par
With the results given in Eq.(25), Ea.(26) and Eq.(64) we completed
the analysis for the decay $\Upsilon\to\gamma+f_2$,
in which the nonperturbative effect related
to $\Upsilon$ and that related to $f_2$ are separated, these effects are
parameterized by NRQCD matrix elements and the distribution amplitudes
$F_0$, $F_1$ and $G_i(i=0,1,2)$, respectively. The distribution
amplitudes characterize how gluons are converted into the tensor meson
$f_2$. Their definitions are given in this and last section. They are
invariant under Lorentz boosts in the $z$-direction, and are gauge
invariant. With the analysis we find that the decay amplitude with
$\lambda =1$ and $\lambda =2$ are suppressed by the power
$\Lambda/m_b$ and $(\Lambda/m_b)^2$ respectively, in comparison
with the decay amplitude with $\lambda =0$.
\par\vskip20pt
\noindent
{\bf 4. Comparison with Experiment}
\par\vskip20pt
In this section we will compare our results with experiment and we will
also discuss the radiative decay into $\eta$. With the decay amplitudes
derived in the last two sections we obtain the decay width:
\begin{equation}
\Gamma(\Upsilon\to\gamma+f_2) =\frac{2}{9}\pi^2Q^2_b\alpha\alpha_s^2(m_b)
 \left(1-\frac{m^2}{4m_b^2}\right)\cdot\frac{1}{m_b^4}
\langle \Upsilon \vert
O_1^\Upsilon (^3S_1) \vert \Upsilon\rangle \sum_{\lambda=0}^2
 \vert T_\lambda \vert^2
\end{equation}
where we used
\begin{equation}
\langle \Upsilon \vert \psi^\dagger \sigma^i \chi \vert 0 \rangle \langle
0|\chi ^{\dagger }\sigma ^j \psi \vert \Upsilon \rangle = \langle\Upsilon \vert
O_1^\Upsilon (^3S_1) \vert \Upsilon\rangle
\varepsilon^i(\varepsilon^j)^\dagger.
\end{equation}
In the above equation $\varepsilon$ is the polarization vector of $\Upsilon$, the
matrix element $\langle\Upsilon \vert
O_1^\Upsilon (^3S_1) \vert \Upsilon\rangle$
is defined in \cite{BBL} and the average over the spin is implied
in the matrix element. As discussed in the last section the relative
order of magnitude of the three amplitudes is given by
\begin{equation}
\vert T_0\vert : \vert T_1 \vert :\vert T_2 \vert
 ={\cal O}(1):{\cal O}(\frac{\Lambda}{m_b}):{\cal O}(\frac{\Lambda^2}{m_b^2}),
\end{equation}
where $\Lambda=\Lambda_{QCD}$ or $k^-$,
and each amplitude receives correction at the order
${\cal O}(\frac{\Lambda^2}{m_b^2})$ relatively to its leading-order contribution.
The corrections are unknown. For consistency we neglect terms with $T_1$ and $T_2$
and the mass $m$ of $f_2$ also  should be neglected. Hence we approximate
the decay width by
\begin{eqnarray}
\Gamma(\Upsilon\to\gamma+f_2) &=& \frac{2}{9}\pi^2Q_b^2\alpha\alpha_s^2(m_b)
 \left(1-\frac{m^2}{4m_b^2}\right)\cdot\frac{1}{m_b^4}
\langle \Upsilon \vert
O_1^\Upsilon (^3S_1) \vert \Upsilon\rangle
 \vert T_0\vert^2
\nonumber\\
&& \cdot \left\{ 1 + {\cal O}(\frac{\Lambda^2}{m_b^2})
  +{\cal O}(\alpha_s ) +{\cal O}(v^2) \right\},
\end{eqnarray}
where the orders of possible corrections are given in $\{\cdots\}$.
For $\Upsilon$ it is a good approximation to neglect these corrections,
because $m_b$ is large, while for $J/\Psi$ the corrections may be significant
with the relatively small $m_c$, each correction in $\{\cdots\}$ can
be at order of $30\%$. We also neglect these for $J/\Psi$.
To compare with experiment we build the ratio
\begin{eqnarray}
 R_h(\Upsilon) &=& \frac{ \Gamma(\Upsilon\to\gamma+f_2)}{
 \Gamma(\Upsilon \to {\rm light\ hadrons})}
\nonumber\\
 &\approx & \frac{27}{10}Q_b^2 \frac{\alpha}{\alpha_s(m_b)} \frac{\pi^2}{\pi^2-9}
 \left(1-\frac{m^2}{4m_b^2}\right) \frac{1}{m_b^2} \vert T_0 \vert^2
\end{eqnarray}
where we used the result at leading orders  for
$ \Gamma(\Upsilon \to {\rm light\ hadrons})$, where
light hadrons are produced through gluon emission. Because Feyman diagrams
for $\Gamma(\Upsilon \to {\rm light\ hadrons})$ have a similar
structure as those given in Fig. 1,  corrections from higher
orders of $\alpha_s$ and of $v^2$ can be cancelled at certain level
in this ratio, we can expect that the ratio receives smaller corrections
than the width does. This ratio does not depend
on the nonperturbative property of $\Upsilon$ in our approximation. The value
of $T_0$ is unknown, it can be extracted from $R_h(\Upsilon)$, and used
to predict $R_h(J/\Psi)$. As an estimation we neglect
the $\mu$-dependence of $T_0$. With
experimental data for hadronic branching ratios we obtain the
ratio:
\begin{eqnarray}
R &=& \frac{ B(\Upsilon\to\gamma +f_2)}{B(J/\Psi\to\gamma +f_2)}
\nonumber\\
 &\approx& \frac{0.923}{0.707}\cdot \frac{Q_b^2m_c^2}{Q_c^2m_b^2}
 \cdot\frac{\alpha_s(m_c)}{\alpha_s(m_b)}\cdot\frac{1-\frac{m^2}{4m_b^2}}
 {1-\frac{m^2}{4m_c^2}}
\nonumber\\
&\approx& 0.059
\end{eqnarray}
where we have used the parameters:
\begin{equation}
m_c=1.5{\rm GeV}, \ \ m_b=5.0{\rm GeV},\ \  \alpha_s(m_c)=0.3,
\ \ \alpha_s(m_b)=0.18.
\end{equation}
Experimentally, the branching ratio has been measured recently, its value
is\cite{CU}:
\begin{equation}
B(\Upsilon\to\gamma +f_2)=(8.1\pm 2.3^{+2.9}_{-2.7})\times 10^{-5}.
\end{equation}
With the value for $B(J/\Psi\to\gamma +f_2)=1.38\times 10^{-3}$
\cite{PDG} we
obtain the ratio determined by experiment:
\begin{equation}
R\approx 0.058\pm 0.016 ^{+0.021}_{-0.020}.
\end{equation}
This value is in agreement with our prediction. However, it should
be kept in mind that experimental errors are large and corrections to
our prediction can be substantial. Another uncertainty is from
the pole-mass of $c$-quark, which value is not well known as the
pole-mass of $b$-quark. A recent QCD sum rule analysis for charmonium
systems\cite{EJa} shows that $m_c$ can be $1.7$GeV. If we vary
the mass $m_c$ from $1.3$GeV to $1.7$GeV, then the ratio $R$ in Eq.(71)
changes from $0.044$ to $0.076$.
\par
As discussed before, we should neglect the decay amplitude with
$\lambda =\pm 1$ and with $\lambda =\pm 2$ to consistently
predict decay widths in our approximation. However the helicity amplitudes
can be measured by measuring the polarization of $f_2$, in which the main
decay mode $f_2 \to\pi\pi$ can be used. This type of measurements have been
done for $J/\Psi$-decay\cite{EJ}. But, as pointed out in \cite{KK}, the fitting
formula used in the measurement for the angular
distribution of $\gamma$ and $\pi$ was not correct.
Hence we still do not know about various helicity amplitudes. With
the sample of $5\times 10^{7}\ J/\psi$'s, whose collection will be ended
this year at BES, a new analysis will provide such information. In our approach,
the angular distribution will be approximated by
\begin{eqnarray}
 \frac{ d{\cal N}}{d\cos\theta_\gamma d\phi_\pi d\cos\theta_\pi}
   &\propto&  \big\{ (1+\cos^2\theta_\gamma)(3\cos^2\theta_\pi-1)^2
 \vert T_0\vert^2
\nonumber\\
&& +{\sqrt 3} \sin 2\theta_\gamma \cos\theta_\pi
 \sin 2\theta_\pi (3\cos^2\theta_\pi-1) {\rm Re}(T_0T_1^*)
 \big\}
 \nonumber\\
&& \cdot \big\{1+{\cal O} (\frac{\Lambda^2}{m_b^2}) \big\}
\end{eqnarray}
where $\theta_\gamma$ is the polar angle between $\gamma$ and the $e^+e^-$-beam
axis, $\theta_\pi$ and $\phi_\pi$ are the polar and azimuthal angles of
a pion in the $f_2$-rest frame with respect to the $\gamma$-direction, and
$\phi_\pi=0$ is defined by the $e^+e^-$-beam axis. A general formula  for
the distribution is provided in \cite{KK}.
\par
With the experimental result in Eq.(73) one can estimate a phenomenological
constant for the gluon conversion into $f_2$. For this we write the
distribution amplitude as
\begin{equation}
 F_0(x)=f_g^S \cdot f(x), \ \ \ {\rm with}\ \int_0^1 f(x) =1
\end{equation}
where the constant $f_g^S$ has dimension 1 in mass and is defined
by\cite{BK}
\begin{eqnarray}
\langle f_2(k)\vert G^a_{\alpha\beta}(0)G^a_{\mu\nu}(0) \vert 0 \rangle
  &=& f_g^T \left\{ \big[ (k_\alpha k_\mu -\frac{1}{2} m^2 g_{\alpha\mu}
  ) \varepsilon^*_{\beta\nu} -(\alpha\leftrightarrow\beta )\big ] -
   (\mu\leftrightarrow\nu) \right\}
\nonumber \\
&&  +\frac{1}{2}f_g^S m^2 \left\{ \big[ g_{\alpha\mu}\varepsilon^*_{\beta\nu}
 -(\alpha\leftrightarrow\beta )\big ] -
   (\mu\leftrightarrow\nu) \right\}.
\end{eqnarray}
If we take the renormalization scale $\mu =m_b$ as a very large scale,
we may take the asymptotic form for the function $f(x)$ at
$\mu =m_b$:
\begin{equation}
f(x) =15x^2(1-x)^2.
\end{equation}
with the experimental data we obtain the estimation at $\mu=m_b$:
\begin{equation}
f_g^S \approx 44{\rm MeV}.
\end{equation}
\par
Before ending this section we briefly discussed the decay
\begin{equation}
 \Upsilon\to \gamma(q) + \eta(k)
\end{equation}
This decay has been studied in \cite{KK}, based on a static quark model for
$\eta$, and the corresponding decay of $J/\psi$ also has been studied
with various approaches\cite{Kuang,Nov,Zhao,FKS}. In these approaches
the decay is controlled by the $U_A(1)$
anomaly, in which the conversion of gluons into $\eta$
is characterized by a local operator of gluons.
In our approach the corresponding operator becomes non-local.
Following the results presented
in Sect. 2 we can obtain the decay amplitude at twist-2 level:
\begin{eqnarray}
\langle \gamma \eta  |S|\Upsilon\rangle &=&
  \frac {-i}{48}eQ_bg_s^2(2\pi )^4\delta^4(2p-k-q)\varepsilon_\rho
\langle 0|\chi ^{\dagger }\sigma ^\ell \psi |\Upsilon  \rangle \
  e^{\ell\rho} \frac{1}{m_b^2}
\nonumber\\
 && \cdot\frac{m^2_\eta}{m_b^2}  \int dx_1 \frac{1-2x_1}{x_1(1-x_1)^2}
  F_\eta (x_1),
\end{eqnarray}
where the tensor $e^{\mu\nu}$ is defined by
\begin{equation}
 e^{\mu\nu}=\varepsilon^{\mu\nu\alpha\beta} l_\alpha n_\beta,
\end{equation}
where $\varepsilon^{\mu\nu\alpha\beta}$ is totally anti-symmetric
with $\varepsilon^{0123}=1$.
The distribution amplitude $F_\eta (x_1)$ characterizes
the conversion of two gluons into $\eta$, which is defined
in the light-cone gauge as
\begin{equation}
F_\eta (x_1)= \frac{1}{2\pi k^+} \int dx^-e^{-ix_1k^+x^-}
\langle \eta (k) \vert G^{a,+\mu}(x^-) G^{a,+\nu} (0) \vert 0\rangle
 e_{\mu\nu}.
\end{equation}
Our result shows that the twist-2 contribution is suppressed by $m_\eta^2$.
In the twist-expansion the light hadron mass $m_\eta$ should be
taken as a small scale as $\Lambda_{QCD}$, hence the contribution
is proportional to $\Lambda^2_{QCD}$. This implies that a complete
analysis should include not only this contribution but also
twist-4 contributions,
in which one needs to consider
the contributions from emission of 2, 3 and 4 gluons.
However, without a complete
analysis we can always write the result of a complete analysis
as
\begin{equation}
\langle \gamma \eta  |S|\Upsilon\rangle =
  \frac {-i}{48}eQ_bg_s^2(2\pi )^4\delta^4(2p-k-q)\varepsilon_\rho
\langle 0|\chi ^{\dagger }\sigma ^\ell \psi |\Upsilon  \rangle \
  e^{\ell\rho} \frac{1}{m_b^4}\cdot g_\eta,
\end{equation}
where the parameter $g_\eta$ has a dimension 3 in mass. The parameter
characterized the conversion of gluons
into $\eta$ and it does not depend on properties of $\Upsilon$.
Hence the above result also applies for the $J/\Psi$ decay. Following
the above analysis for $f_2$ we obtain for $\eta$:
\begin{eqnarray}
R &=& \frac{ B(\Upsilon\to\gamma +\eta)}{B(J/\Psi\to\gamma +\eta)}
\nonumber\\
 &\approx& \frac{0.923}{0.707}\cdot \frac{Q_b^2m_c^6}{Q_c^2m_b^6}
 \cdot\frac{\alpha_s(m_c)}{\alpha_s(m_b)}\cdot\frac{1-\frac{m_\eta^2}{4m_b^2}}
 {1-\frac{m_\eta^2}{4m_c^2}}
\nonumber\\
 &\approx & 4.1\times 10^{-4}.
\end{eqnarray}
With the experimental data for $B(J/\Psi\to\gamma +\eta)$ we obtain
\begin{equation}
B(\Upsilon\to\gamma +\eta)\approx 3.5\times 10^{-7}.
\end{equation}
The predicted $B(\Upsilon\to\gamma +\eta)$ may be too small to be observed.
\par\vskip20pt
\noindent
{\bf 5. Conclusion}
\par\vskip20pt
In this work we have studied the radiative decay of a $^3S_1$ quarkonium into
a tensor meson $f_2$. A factorization at tree-level is performed for the decay
amplitude, in which nonperturbative effect related to the quarkonium and
that related to $f_2$ is separated. These effects are parameterized with
NRQCD matrix elements and with a set of gluonic distribution amplitudes,
which  are defined in this work. These amplitudes are gauge invariant
and universal.
At twist-2 level, we find two amplitudes characterizing the conversion
of two gluons into $f_2$ with the helicity $\lambda =0$ and $\lambda =2$,
respectively, and the contribution to the decay amplitude with $\lambda =2$
is suppressed by $m^2$. At twist-3 level the decay amplitude is nonzero
only for $\lambda =1$, three gluonic distribution amplitudes
are introduced to describe the gluon conversion. At loops-level, the
factorization may still be performed as shown in studies of other
exclusive processes\cite{Collins}, but we will not study this subject
in this work
and leave it for future.
\par
In our approach an expansion in $v$, the velocity of the $b$-quark inside
$\Upsilon$ in its rest-frame, is used. We have only taken the leading order
contributions at $v^0$. At this order, $\Upsilon$ can be considered as a
bound state of the $b\bar{b}$ pair in a color singlet. The corrections to the
leading-order results may also added. However, problems arise at order of $%
v^4$. At this order $\Upsilon$ has a component in which the $b\bar{b}$ pair
is in color octet and the component is a bound state of the $b\bar{b}$ pair
with soft gluons. It is unclear how to add corrections from this component
at order of $v^4$. It deserves a further study of these problems to
understand a bound state of many dynamical freedoms of QCD.
\par
The NRQCD matrix elements are known from other
experiment and form lattice QCD calculations, while there is not
many information about the 5 gluonic distribution amplitudes.
Using experiment data and our result we extract a parameter relevant
for the $\lambda =0$ amplitude. It should be emphasized that these
distribution amplitudes are universal and can be used for production
of $f_2$ in other processes, like the one studied in \cite{BK}.
In experiment, a measurement of the polarization of $f_2$
can provide information about these distribution amplitudes with
different $\lambda$.  With
the sample of $5\times 10^{7}\ J/\psi$'s, whose collection will be ended
this year at BES, such measurement is planed\cite{Zhu}.
\par
With our results we can predict the ratio of the branching ratios for
$\Upsilon$ and for $J/\Psi$. An agreement between the prediction and
experiment can be obtained. In the approach used here, we can also
perform an analysis for the radiative decay into $\eta$. It turns out
that twist-4 effect should be included in a complete analysis. But
without the complete analysis we are still able to predict
the branching ratio for $\Upsilon$ and it may be too small to be observed.
\par

\vskip20pt \noindent
{\bf Acknowledgment:}  This work
is supported by National Science Foundation of P.R. China and by the Hundred
Young Scientist Program of Sinica Academia of P.R.China.

\vskip15pt

\vfil\eject

\vfil\eject

%\vskip 10mm

\begin{center}
\begin{picture}(200,60)(0,0)
%\SetScale{0.5}
\SetWidth{1.0}

\ArrowLine(0,30)(60,30)
\Text(30,10)[]{$\Upsilon$} \GOval(60,30)(30,10)(0){0.5}

\Line(60,60)(120,60) \Line(60,0)(120,0) \Line(120,0)(120,60)

\Photon(120,0)(200,0){3}{8}

\Gluon(120,60)(160,60){3}{4} \Gluon(120,30)(160,30){3}{4}
\ArrowLine(160,45)(200,45) \GOval(160,45)(15,5)(0){0.5}

\Text(180,55)[b]{$f_2$}

\end{picture}
\vskip 10mm
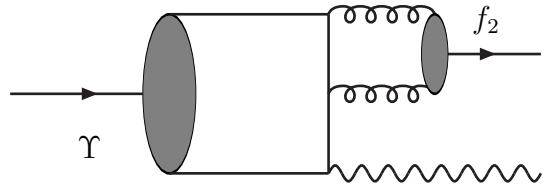
\centerline{Fig.1}
{\bf Fig.1}:  One of the diagrams for contributions of two-gluon emission.
\vfill\eject\pagestyle{plain}
\end{center}

\begin{center}
\begin{picture}(200,150)(20,0)
%\SetScale{0.5}

\SetWidth{1.0}

\ArrowLine(20,150)(60,150)
\Text(40,130)[]{$\Upsilon$} \GOval(60,150)(30,10)(0){0.5}

\Line(60,180)(90,180) \Line(60,120)(90,120) \Line(90,120)(90,180)

\Photon(90,120)(200,120){2}{8}

\Gluon(90,180)(160,180){2}{8} \Gluon(90,150)(160,150){2}{8}
\Gluon(90,165)(160,165){2}{8} \ArrowLine(160,165)(200,165)
\GOval(160,165)(15,5)(0){0.5}

\Text(180,175)[b]{$f_2$} \Text(110,100)[]{(a)}

\ArrowLine(20,30)(60,30)
\Text(40,10)[]{$\Upsilon$} \GOval(60,30)(30,10)(0){0.5}

\Line(60,60)(90,60) \Line(60,0)(90,0) \Line(90,0)(90,60)

\Photon(90,0)(200,0){2}{8}

\Gluon(90,60)(125,60){-2}{4} \Gluon(125,60)(160,60){-2}{4}
\Gluon(125,58)(160,45){2}{4} \Vertex(125,59){3}

\Gluon(90,30)(160,30){2}{8} \ArrowLine(160,45)(200,45)
\GOval(160,45)(15,5)(0){0.5}

\Text(180,55)[b]{$f_2$} \Text(110,-20)[]{(b)}

\end{picture}
\vskip 10mm
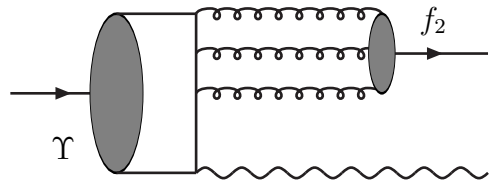
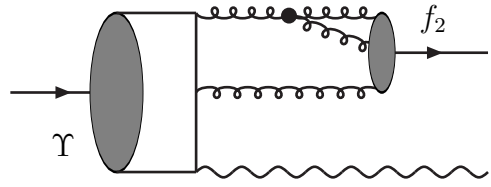
\centerline{Fig.2}
{\bf Fig.2}: Typical diagrams for contributions with three-gluon emission.
%\vfill\eject\pagestyle{plain}\setcounter{page}{0}
\end{center}
\end{document}